\documentclass[prl,twocolumn]{revtex4}
\usepackage{graphicx}
\usepackage{dcolumn}
\usepackage{color}
\usepackage{latexsym}
\usepackage{bm}
\begin{document}

\title{Hawking radiation and near horizon universality of 
chiral Virasoro algebra }
\author{Rabin Banerjee}
\email{rabin@bose.res.in}
\author{Sunandan Gangopadhyay}
\email{sunandan@bose.res.in}
\author{Shailesh Kulkarni}
\email {shailesh@bose.res.in}
\affiliation{
Satyendra Nath Bose National Centre for Basic Sciences, Block-JD, 
Sector-III, Salt Lake, Kolkata 700 098, India
}
\begin{abstract}
\noindent We show that the diffeomorphism anomaly together with
the trace anomaly reveal a chiral Virasoro algebra 
near the event horizon of a black hole. 
This algebra is the same irrespective of whether the anomaly
is covariant or consistent, thereby manifesting its universal
character and the fact that only the outgoing modes are relevant
near the horizon. Our analysis therefore clarifies the role of the
trace anomaly in the diffeomorphism anomaly approach 
\cite{wilczek, isowilczek, shailesh, 
shailesh2, sunandan, sunandan10, rabin10} 
to the Hawking radiation.

\end{abstract}

\maketitle

\noindent The diffeomorphism anomaly cancellation approach to
Hawking radiation \cite{wilczek, isowilczek, shailesh, shailesh2, 
sunandan, sunandan10, rabin10} 
has been the centre of much attraction recently. 
The basic idea is that the effective theory near the horizon of a black hole 
becomes two dimensional and chiral. Hence it is anomalous.
An anomaly signifies the breakdown of a classical 
symmetry upon the process of quantisation. In this case there is a violation 
of classical general coordinate invariance 
leading to a nonconservation of the energy momentum tensor, 
called the diffeomorphism anomaly. This diffeomorphism anomaly
admits two different forms -- the consistent and the covariant -- which
are related by local counterterms \cite{bert}.
Using either of the forms of the two dimensional 
gravitational anomaly, Hawking fluxes
are obtained. This can be compared to the derivation
of Christensen and Fulling \cite{christ}, where they show that
far away from the horizon the trace anomaly, along with the conservation
of energy momentum tensor (i.e. absence of any diffeomorphism anomaly), 
are required to give the Hawking flux in two dimensions.
However, the trace anomaly near the horizon 
in the diffeomorphism anomaly method 
does not play any role in deriving the Hawking flux.

\noindent In this paper, we observe that 
although the trace anomaly near the horizon 
does not play a role in deriving the Hawking flux 
in the diffeomorphism anomaly method, it is important in its
own right. The diffeomorphism anomaly together with the trace anomaly
near the horizon lead to a chiral Virasoro algebra 
of matter fields near the horizon
which reassures the fact that it is only the outgoing modes which are
relevant near the horizon. An important point to note
is that identical results are obtained using either the 
consistent or covariant forms of the above anomalies,
thereby establishing the universality
of the chiral Virasoro algebra near the horizon.

\noindent To appreciate this aspect, we observe that the results are  
derived by using yet another manifestation of the 
diffeomorphism as well as the trace anomaly
which modifies the standard algebra of commutators involving
components of energy momentum tensor. This manifestation
is different for the consistent and covariant anomalies 
(as we shall see below) but finally leads to the same chiral Virasoro algebra 
of matter fields near the horizon. We shall first consider the case
of the consistent anomaly and then the covariant anomaly.
   
\noindent We start from the metric of the static spherically 
symmetric black hole  
\begin{equation}
ds^2 = f(r) dt^2 - \frac{1}{f(r)}dr^2 -r^2d\Omega^2
\label{1}
\end{equation}
where $f(r=r_H)=0$, with $r=r_H$ being the event horizon of the black hole.
As we have already mentioned, the theory near the horizon 
is effectively two dimensional, 
with the metric given by the $r-t$ sector
of (\ref{1}). The ingoing (left-handed) modes 
traversing inside the black hole are lost. 
Hence, near the horizon, the two dimensional effective theory becomes chiral
comprising only the outgoing (right-handed) modes.  
Such a theory has a diffeomorphism anomaly whose consistent
form is given by \cite{Leut, Bardeen, bertlmann}
\begin{equation}
\nabla^{\mu}T_{\mu\nu} = - \frac{c_{R} - c_{L}}{96\pi} 
\epsilon^{\gamma\delta}
\partial_{\alpha}\partial_{\gamma}{\Gamma^{\alpha}}_{\delta\gamma} 
= \mathcal{A}_{\nu}~.
\label{1aa}  
\end{equation}
Correspondingly, there is a  consistent Weyl (non-vanishing trace) anomaly
\begin{equation}
{T^{\mu}}_{\mu} = \frac{1}{48\pi}(c_{L} + c_{R})(R-\frac{1}{2}
\epsilon^{ab}\nabla_{\mu}\omega^{\mu}_{ab})
\label{2aa}    
\end{equation}
where, $\omega^{a}_{b\mu}$ are the spin connections given by
\begin{equation}
\omega^{a}_{b\mu}=e^{a}_{\nu}\nabla_{\mu}
E^{\nu}_{b}=e^{a}_{\nu}
\partial_{\mu}E^{\nu}_{b} + e^{a}_{\nu}
\Gamma^{\nu}_{\nu\lambda}E^{\lambda}_{b}
\label{spinconn}    
\end{equation}
and $e^{a}_{\mu}$ is the zweibein and $E^{\mu}_{a}$ 
its inverse $E^{\mu}_{a}e^{a}_{\nu} =\delta^{\mu}_{\nu}$.
$c_{L}$ and $c_{R}$ denote the central charges of the left 
and right moving modes respectively and $R$ is the Ricci scalar 
corresponding to the $r-t$ sector of the metric.
For a vector theory $c_{L}=c_{R}$ so that there is no diffeomorphism anomaly
and the energy momentum tensor is 
covariantly conserved $(\nabla_{\mu}{T^{\mu}}_{\nu} =0)$.
Only the Weyl anomaly exists. 
This is the scenario away from the horizon from which, following
\cite{christ}, the flux may be determined. Here, on the contrary,
the theory is chiral and we set $c_{L} = 0$ so that
both the anomalies are present. Henceforth we always take $c_{L}=0$. 
\noindent The consistent form of the Weyl anomaly (\ref{2aa}) 
leads \cite{tomiya} to the following commutator 
among the components of the consistent energy momentum tensor
\begin{eqnarray}
[T_{00}(x), T_{01}(x')] &=& i[T_{00}(x) + T_{00}(x')]
\partial_{x}\delta(x-x')\nonumber\\
&& +\frac{i c_{R}}{24\pi}\partial_{x}^3\delta(x-x')~.\label{Y}
\end{eqnarray}
The absence of a trace anomaly would yield a vanishing Schwinger term
appearing in (\ref{Y}).

\noindent Likewise, the consistent form of the diffeomorphism anomaly 
(\ref{1aa}) yields \cite{tomiya}      
\begin{eqnarray}
[T_{01}(x),T_{01}(x')] &=& i[T_{01}(x)+T_{01}(x')]
\partial_{x}\delta(x-x')\nonumber\\
&& + \frac{i c_{R}}{24\pi}\partial_{x}^{3}\delta(x-x') \label{X}
\end{eqnarray}
\begin{eqnarray}
[T_{00}(x),T_{00}(x')]&=& i[T_{01}(x)+T_{01}(x')]
\partial_{x}\delta(x-x') \nonumber\\
&& + \frac{i c_{R}}{24\pi}\partial_{x}^{3}\delta(x-x')~.\label{X'}
\end{eqnarray}
Indeed if there was no diffeomorphism anomaly (\ref{1aa}),
the Schwinger terms appearing in  (\ref{X}) and (\ref{X'}) 
would be absent \cite{schwinger}.\\
\noindent The chiral nature becomes transparent by
introducing the light cone combinations 
\begin{eqnarray}
T_{00} &=& T_{++} + T_{--}\nonumber \\
T_{01} &=& T_{++} - T_{--}~. 
\label{3}
\end{eqnarray}  
Using (\ref{Y}), (\ref{X}) and (\ref{X'}),
we obtain the following non-vanishing algebra 
among the light cone components of $T_{\mu\nu}$
\begin{eqnarray}
[T_{++}(x), T_{++}(x')] &=& i[T_{++}(x)+T_{++}(x')]
\partial_{x}\delta(x-x')\nonumber\\
&& + \frac{i c_{R}}{24\pi} \partial_{x}^{3}\delta(x - x')
\label{400}
\end{eqnarray}
while the algebra of $T_{--}$ is trivial with no Schwinger
terms. The quantum effect is totally encoded in $T_{++}$. 
This allows us to put $T_{--} = 0$ and proceed 
with only $T_{++}$.\\

\noindent Using these anomalous commutators, it is possible to
compute the chiral Virasoro algebra which is a new ingredient of our paper.
To do this, the chiral Virasoro generator is defined as
\begin{equation}
L_{n} = \frac{1}{2}\int_{0}^{\pi} dx ~T_{++}(x)~e^{2i n x} . \label{5}  
\end{equation}
Using (\ref{400}), the commutator 
between $L_{n}$ and $L_{m}$ yields 
\begin{equation}
[L_{n},L_{m}] = (n-m)L_{n+m} 
+\frac{c_{R}}{12}m^3\delta_{n+m,0}~. \label{6}
\end{equation}
The above result shows that we have a chiral Virasoro
algebra of matter fields near the horizon reassuring the fact that it
is indeed the outgoing modes which are relevant near the
horizon. We also observe that the central charge $c_{R}$ introduced in the
diffeomorphism anomaly (\ref{1aa}) yields the appropriate
central extension in the chiral Virasoro algebra with the 
correct normalisation. \\  
\noindent We now show that the same chiral Virasoro algebra is once again
recovered if we work with the covariant form of the anomalies near the horizon.
The covariant form of the diffeomorphism anomaly reads
\cite{Leut, Bardeen, bertlmann}
\begin{equation}
\nabla^{\mu}\tilde{T}_{\mu\nu} = - \frac{c_{R} - c_{L}}{96\pi} 
\epsilon_{\mu\nu}\nabla^{\mu}R=\tilde{A}_{\nu}
\label{covdiff}  
\end{equation}
and the covariant form of the Weyl (non-vanishing trace) anomaly reads
\begin{equation}
{\tilde{T}^{\mu}}_{\mu} = \frac{1}{48\pi}(c_{L} + c_{R})R~.
\label{covtr}    
\end{equation}
To find the manifestation of the above covariant anomalies in terms
of the commutators of the covariant energy  momentum tensor, a little
amount of work is needed. Note that the covariant and the consistent 
energy  momentum tensors are related by local counter terms :
\begin{eqnarray}
\tilde{T}_{\mu\nu} = T_{\mu\nu}+P_{\mu\nu}
\label{local}    
\end{eqnarray}
where, 
\begin{eqnarray}
\nabla^{\mu}P_{\mu\nu}=-\frac{1}{96\pi}
(\epsilon_{\mu\nu}\nabla^{\mu}R-\epsilon^{\gamma\delta}\partial_{\alpha}
\partial_{\gamma}{\Gamma^{\alpha}}_{\delta\nu}).
\label{gradlocal}    
\end{eqnarray}  
The forms of the local counter terms can be computed easily
for the static spherically symmetric black hole spacetime 
(\ref{1}) by solving the
above equation and reads
\begin{eqnarray}
{P^{r}}_{t}=\frac{1}{192\pi}[ff''-2f'^2]\quad;\quad {P^{t}}_{t}={P^{r}}_{r}=0.
\label{gradlocal}    
\end{eqnarray} 
Hence the covariant forms of the commutators 
(\ref{Y}), (\ref{X}) and (\ref{X'})read 
\begin{eqnarray}
[\tilde T_{00}(x), \tilde T_{01}(x')] &=& 
i([\tilde T_{00}(x) + \tilde T_{00}(x')] - [P_{00}(x) + P_{00}(x')])
\nonumber\\
&& \times\partial_{x}\delta(x-x')
+\frac{i c_{R}}{24\pi}\partial_{x}^3\delta(x-x')~.\label{Y''}
\end{eqnarray}
\begin{eqnarray}
[\tilde T_{01}(x),\tilde T_{01}(x')] &=& 
i([\tilde T_{01}(x)+\tilde T_{01}(x')]-[P_{01}(x)+P_{01}(x')])\nonumber\\
&&\times\partial_{x}\delta(x-x')+\frac{i c_{R}}{24\pi}\partial_{x}^{3}
\delta(x-x') \label{X''}
\end{eqnarray}
and 
\begin{eqnarray}
[\tilde T_{00}(x),\tilde T_{00}(x')]&=& 
i([\tilde T_{01}(x)+\tilde T_{01}(x')]-[P_{01}(x)+P_{01}(x')])\nonumber\\
&&\times\partial_{x}\delta(x-x')+ 
\frac{i c_{R}}{24\pi}\partial_{x}^{3}\delta(x-x')\label{z''}
\end{eqnarray}
where,
\begin{eqnarray}
P_{01}&=&P_{10}\equiv P_{rt}=-\frac{1}{f}{P^{r}}_{t} \nonumber\\ 
P_{00}\equiv P_{tt}&=& f{P^{t}}_{t}= 0~. \label{A''}
\end{eqnarray} 
Using these commutation relations and introducing the light cone combinations
of the covariant energy momentum tensor as before, we get the algebra among
the light cone components of covariant energy-momentum tensor
\begin{eqnarray}
[\tilde T_{++}(x), \tilde T_{++}(x')] &=& 
i([\tilde T_{++}(x)+\tilde T_{++}(x')]\nonumber\\
&&-[P_{++}(x)+P_{++}(x')])\partial_{x}\delta(x-x')\nonumber\\ 
&&+ \frac{i c_{R}}{24\pi} \partial_{x}^{3}\delta(x - x')
\label{400'}
\end{eqnarray}
where, 
\begin{eqnarray}
P_{++}(x) &=& \frac{1}{2}[P_{00}(x)+P_{01}(x)]=-\frac{1}{2f}{P^{r}}_{t} 
\label{400'''}
\end{eqnarray}
and $\tilde T_{++}$, $\tilde T_{--}$ are the covariant 
analogues of (\ref{3}). Note that the algebra among $\tilde T_{--}$
is trivial as in the case of consistent anomaly revealing the chiral nature
of the theory once again. \\
Next, we define the covariant chiral Virasoro generator 
\begin{equation}
\tilde L_{n} = \frac{1}{2}\int_{0}^{\pi} dx ~\tilde T_{++}(x)~e^{2i n x} . 
\label{5'}  
\end{equation}
Using (\ref{400'}), the commutator among $\tilde L_{n}$ and $\tilde L_{m}$
yields
\begin{eqnarray}
[\tilde L_{n},\tilde L_{m}] &=& (n-m)(\tilde L_{n+m} - G(n,m))\nonumber\\ 
&&+\frac{c_{R}}{12}m^3\delta_{n+m,0} \label{6'}
\end{eqnarray} 
where the function $G(n,m)$ is given by
\begin{equation}
G(n,m) = 2\int_{0}^{\pi} dx ~ e^{2i(m+n)x}P_{++}(x)~.\nonumber
\end{equation}
Now, by redefining the Virasoro generator as
\begin{equation}
\tilde L_{n} \rightarrow \tilde L'_{n} = \tilde L_{n} - G(n) 
\label{modvirasoro} 
\end{equation}
the Virasoro algebra (\ref{6}) is once again recovered
\begin{equation}
[\tilde{L}^{'}_{n},\tilde{L}^{'}_{m}] = (n-m)\tilde{L}^{'}_{n+m} 
+\frac{c_{R}}{12}m^3\delta_{n+m,0}~. \label{covvira}
\end{equation}
Thus we observe that although the Schwinger terms in the anomalous
commutators (\ref{400}) and (\ref{400'}) are different for the consistent
and covariant cases, both lead to the same chiral Virasoro algebra.\\

\noindent We finally conclude by making some observations.
It should be noted that the chiral Virasoro
algebra near the horizon has also been obtained in the context of classical 
Liouville theory \cite{solodukhin} which is eventually
used in the computation of the black hole entropy.
Interestingly, the chiral Virasoro algebra obtained in \cite{solodukhin}
has the same functional form as found here in (\ref{6}, \ref{covvira}).
We therefore feel that the ideas of deriving the Hawking radiation from
diffeomorphism anomaly of an effective two dimensional chiral theory of quantum
fields in a background black hole spacetime can 
also be used in the abstraction of black hole entropy. A proper
understanding of this connection may give 
us an important insight about quantum black holes.


\end{document}